\newcommand{\p}{\mathcal{P}}
\newcommand{\q}{\mathcal{Q}}

\documentclass[12pt]{article}
\usepackage{amsmath}
\usepackage{amssymb}
\usepackage{amscd}
\usepackage[dvips]{graphicx}
\begin{document}
~~
\bigskip
\bigskip
\bigskip
\bigskip
\bigskip

\begin{center}

\section*{$\kappa$-DEFORMED OSCILLATORS, THE CHOICE OF STAR PRODUCT AND FREE $\kappa$-DEFORMED QUANTUM FIELDS}
\end{center}
\bigskip
\bigskip
\bigskip
\begin{center}
{{\large\bf ${\rm Marcin\;Daszkiewicz}$, ${\rm Jerzy\; Lukierski}$,
${\rm Mariusz\;Woronowicz}$}}
\end{center}
\begin{center}
{%\large
{Institute of Theoretical Physics\\ University of Wroc{\l}aw pl.
Maxa Borna 9, 50-206 Wroc{\l}aw, Poland\\ e-mail:
marcin,lukier,woronow@ift.uni.wroc.pl}}
\end{center}
\bigskip
\bigskip
\bigskip
\bigskip
\bigskip
\bigskip

\begin{abstract}
The aim of this paper is to study the general framework providing
various  $\kappa$-deforma-\\tions of field oscillators  and consider
the commutator function of the corresponding $\kappa$-deformed free
fields. In order to obtain free $\kappa$-deformed quantum fields
(with $c$-number commutators) we proposed earlier a particular model
of $\kappa$-deformed oscillator algebra \cite{11} and the
modification of $\kappa$-star product \cite{1}, implementing  in the
product of two quantum fields the change of standard
$\kappa$-deformed mass-shell conditions. We recall here that other
different  models of $\kappa$-deformed oscillators recently
introduced in \cite{12}-\cite{ny}  are defined  on standard
$\kappa$-deformed mass-shell. In this paper we consider the most
general $\kappa$-deformed field oscillators, parametrized by set of
arbitrary functions in three-momentum space.  Firstly, we study the
fields with the $\kappa$-deformed oscillators defined  on standard
$\kappa$-deformed mass-shell, and  argue that for any such a choice
of  $\kappa$-deformed  field oscillators algebra   we do not obtain
 the free quantum $\kappa$-deformed fields with
the $c$-number commutators. Further, we study $\kappa$-deformed
quantum fields with the modified $\kappa$-star product and derive
large class of $\kappa$-oscillators defined on suitably modified
 $\kappa$-deformed mass-shell. We obtain
large class of $\kappa$-deformed statistics depending on six
arbitrary functions which all provide the $c$-number field
commutator functions. This  general class of $\kappa$-oscillators
can be described by the composition of suitably defined
 $\kappa$-multiplications and   the
$\kappa$-deformation of the flip operator.
\end{abstract}
\bigskip
\bigskip
\bigskip
\bigskip

\eject

\section{Introduction}

The standard relativistic local quantum fields on Minkowski space
provide a basic tool for the description of fundamental
interactions. If we include into consideration the quantum gravity
effects such classical description breaks down at Planck distances
($\approx 10^{-33}$ cm) and it is quite plausible that space-time
becomes noncomutative (see e.g. \cite{2}, \cite{3}). The
$\kappa$-deformation of Minkowski space \cite{5}-\cite{7} and
corresponding quantum $\kappa$-deformation of relativistic
symmetries (see e.g. \cite{8}, \cite{9}, \cite{6}) provides a
fundamental mass scale and possible tools for the description of
Planckian quantum-gravitational regime.

At present an important task is the construction of
$\kappa$-deformed quantum field theory on $\kappa$-deformed
Minkowski space. In contrast with simpler, recently studied
 case of $\theta$-deformed symmetries (with constant commutator $[\;{\hat
x}_{\mu},{\hat x}_{\nu}\;] = \theta_{\mu\nu}$; see e.g. \cite{chai},
\cite{wess}), the $\kappa$-deformation of relativistic symmetries is
not described by a twist factor\footnote{We restrict here ourselves
to twists $T\in \mathcal{U}(P_4)\otimes \mathcal{U}(P_4)$ where
$\mathcal{U}(P_4)$ describes the enveloping $D=4$ Poincare algebra.
For twists not satisfying this condition see \cite{azja},
\cite{go}.}, and for the $\kappa$-Poincare algebra the universal
R-matrix is not known. We can therefore apply mainly the technique
of star product (for application to $\kappa$-deformation see e.g.
\cite{czechy}-\cite{meljanac}) as representing the
$\kappa$-deformation of space-time.

a) {\it  Summary of previous results  \cite{11}, \cite{1},
\cite{10}}. Main aim of our scheme presented earlier in \cite{11},
\cite{1}, \cite{10} was the construction of free quantum
$\kappa$-deformed fields characterized by $c$-number commutator
function\footnote{We extend to noncommutative quantum fields the
known definition of (generalized) free fields as described by a
$c$-number commutator (see e.g. \cite{cnumber1}, \cite{cnumber2}).}.
It should be stressed that  for such free $\kappa$-deformed quantum
fields all other properties known from the standard free quantum
field theory as the notion of locality, microcausality, the
structure of Fock space, or kinematic independence of field
excitations defining the multiparticle states are  modified. The
$c$-number commutator of quantum $\kappa$-deformed fields is
obtained by the interplay of the two following sources of
noncommutativity: quantum nature  of space-time and specific
$\kappa$-deformation of the field oscillators algebra:

i) {\it Noncommutativity of space-time}. We replace the standard
quantum field arguments $x_\mu$ by $\kappa$-Minkowski noncommutative
coordinates ${\hat x}_{\mu}$
\begin{equation}
[\;{\hat x}_{0},{\hat x}_{i}\;] = \frac{i}{\kappa}{\hat
x}_{i}\;\;,\;\; [\;{\hat x}_{i},{\hat x}_{j}\;] = 0\;.
\label{minkowski}
\end{equation}
One can  introduce the Fourier expansion of $\kappa$-deformed free
quantum fields
\begin{equation}
{\hat \varphi} ({\hat x}) = \frac{1}{(2\pi)^{\frac{3}{2}}} \int
d^4p\; {\hat A}(p_0,\vec{p})\;\delta \left(C^2_{\kappa}
(\vec{p},p_0) - M^2\right)\vdots {\rm e}^{ip_\mu
\hat{x}^\mu}\vdots\;, \label{field}
\end{equation}
where the symmetrized $\kappa$-deformed plane wave looks as follows
(see e.g. \cite{star1})
\begin{equation}
\vdots {\rm e}^{ip_\mu \hat{x}^\mu}\vdots ={\rm
e}^{\frac{i}{2}p_0{\hat x}_{0}}{\rm e}^{ip_i{\hat x}_{i}} {\rm
e}^{\frac{i}{2}p_0{\hat x}_{0}}\;. \label{exp}
\end{equation}
The operators ${\hat A}(p_0,\vec{p})$ describe the  quantized field
oscillators and $C_2^{\kappa} (\vec{p},p_0)$ represents  the
$\kappa$-deformed mass square Casimir
\begin{equation}
C_2^{\kappa} (\vec{p},p_0) = \left (2\kappa\sinh
\left(\frac{p_0}{2\kappa}\right)\right)^2 - \vec{p}^{\ 2}\;,
\label{casimir}
\end{equation}
defining the $\kappa$-deformed mass-shell condition
\begin{equation}
C_2^{\kappa} (\vec{p},p_0) - M^2 = 0\;, \label{casimirmilion}
\end{equation}
which implies the energy-momentum dispersion relation
\begin{equation}
p_0 = \pm \omega_{\kappa}(\vec{p}) \;,\label{czartworomega}
\end{equation}
where
\begin{equation}
\omega_{\kappa}(\vec{p}) = 2\kappa {\rm
arcsinh}\left(\frac{\omega(\vec{p})}{2\kappa}\right)\;\;\;;\;\;\;\omega(\vec{p})
= \sqrt{\vec{p}^{\ 2} + M^2} \;.\label{omega}
\end{equation}

For the discussion of $\kappa$-deformed free fields we introduce the
operator-valued Weyl homomorphism
\begin{equation}
{\hat \varphi} ({\hat x}) \leftrightarrow {\hat \varphi}
({x})\;\;,\;\;{\hat \varphi} ({\hat y}) \leftrightarrow {\hat
\varphi}({y}) \;, \label{map}
\end{equation}
where the relations (\ref{minkowski}) describing the pair of
noncommutative $\kappa$-Minkowski points ${\hat x}$, ${\hat y}$ is
supplemented by the following additional crossrelations\footnote{The
relation (\ref{doubleminkowski}) permits to perform in consistency
with (\ref{minkowski}) the limit ${\hat x}_{\mu} \to {\hat
y}_{\mu}$. Other option for constructing the $\kappa$-deformed field
theory is to assume that $[\;{\hat x}_{\mu},{\hat y}_{\nu}\;] = 0$
(see e.g. \cite{bdf}).}
\begin{equation}
[\;{\hat x}_{0},{\hat y}_{i}\;] = \frac{i}{\kappa}{\hat
y}_{i}\;\;,\;\; [\;{\hat y}_{0},{\hat x}_{i}\;] =
\frac{i}{\kappa}{\hat x}_{i}\;. \label{doubleminkowski}
\end{equation}
 The corresponding $\kappa$-star multiplication prescription
which represents  the noncommutative space-time structure
(\ref{minkowski}), (\ref{doubleminkowski}) looks as follows
\begin{eqnarray}
&&{\hat \varphi}(\hat{x})\cdot {\hat
\varphi}(\hat{y})\leftrightarrow {\hat \varphi}({x})\star_{\kappa}
{\hat \varphi}({y})= \frac{1}{(2\pi)^{3}} \int d^4p\int d^4q\; {\rm
e}^{i(p_0x_0+q_0y_0) + (p_i{\rm e}^{\frac{q_0}{2\kappa}}x^i +
q_i{\rm
e}^{-\frac{p_0}{2\kappa}}y^i)}\label{newstar1} \\
&&~~~~~~~~~~~~~~~~~~~~~~~~~~~~~~~~~~~~\cdot\; {\hat
A}(p_0,\vec{p}){\hat A}(q_0,\vec{q}) \delta ( C^2_{\kappa}
(\vec{p},p_0) - M^2)~ \delta ( C^2_{\kappa} (\vec{q},q_0) - M^2
)\,.\nonumber
\end{eqnarray}
The formula (\ref{newstar1}) follows from the $\kappa$-star products
of exponentials (\ref{exp})
\begin{equation}
\vdots {\rm e}^{ip_\mu\hat{x}^\mu}\vdots \cdot\vdots {\rm
e}^{iq_\mu\hat{y}^\mu}\vdots \leftrightarrow{\rm e}^{ip_\mu{x}^\mu}
\star_{\kappa} {\rm e}^{iq_\mu{y}^\mu} = {\rm e}^{i(p_0x^0+q_0y^0) +
(p_i{\rm e}^{\frac{q_0}{2\kappa}}x^i + q_i{\rm
e}^{-\frac{p_0}{2\kappa}}y^i)}\;, \label{twopoint}
\end{equation}
where rhs of (\ref{twopoint}) is determined by the fourmomentum
coproduct
\begin{equation}
\Delta(P_0) = P_0\otimes 1 + 1\otimes P_0\;\;\;,\;\;\; \Delta(P_i) =
P_i\otimes {\rm e}^{\frac{P_0}{2\kappa}} + {\rm
e}^{-\frac{P_0}{2\kappa}}\otimes P_i\;.\label{c4}
\end{equation}
The field oscillators in the product
$\hat{\varphi}(x)\star_{\kappa}\hat{\varphi}(y)$ (see
(\ref{newstar1})) carry the fourmomenta satisfying  the standard
$\kappa$-deformed mass-shell condition (\ref{casimirmilion}). We
recall that in the models of $\kappa$-statistics presented in
\cite{12}-\cite{ny} there is imposed such a standard
$\kappa$-deformed mass-shell condition.

In our papers  \cite{1}, \cite{10} we modified  the standard
$\kappa$-star product (\ref{newstar1}) by the following additional
deformation of the product of mass-shell deltas
\begin{equation}
\delta( C^2_{\kappa} (\vec{p},p_0) - M^2)~\delta ( C^2_{\kappa}
(\vec{q},q_0) - M^2 ) \Rightarrow \delta ( C^2_{\kappa} (\vec{p}{\rm
e}^{\frac{q_0}{2\kappa}},p_0) - M^2)~ \delta ( C^2_{\kappa}
(\vec{q}{\rm e}^{-\frac{p_0}{2\kappa}},q_0) - M^2
)\;.\label{slawion}
\end{equation}

Such modification can be also introduced as the change of standard
multiplication of the fields on noncomutative space-time
\begin{eqnarray}
&&{\hat \varphi}(\hat{x})\cdot{\hat \varphi}(\hat{y})\rightarrow
{\hat \varphi}({\hat{x}})\cdot_\kappa{\hat \varphi}({\hat{y}})
=\frac{1}{(2\pi)^{3}} \int d^4p\int d^4q\;
 {\hat A}(p_0,\vec{p}){\hat
A}(q_0,\vec{q}){\rm e}^{i(p\hat{x}+q\hat{y})}\label{slawion1w}\\
&&~~~~~~~~~~~~~~~~~~~~~~~~~~~~~~~~~~~~~~~~~~~~\cdot\;\delta (
C^2_{\kappa} (\vec{p}{\rm e}^{\frac{q_0}{2\kappa}},p_0) - M^2)~
\delta ( C^2_{\kappa} (\vec{q}{\rm e}^{-\frac{p_0}{2\kappa}},q_0) -
M^2 )\,, \nonumber
\end{eqnarray}
which is homomorphic to the following modified $\kappa$-star product
of two $\kappa$-deformed free fields on commuting Minkowski space
\begin{eqnarray}
&&{\hat \varphi}({x}) {\hat \star}_{\kappa} {\hat \varphi}({y}) =
\frac{1}{(2\pi)^{3}} \int d^4p\int d^4q\;
 {\hat A}(p_0,\vec{p}){\hat
A}(q_0,\vec{q}){\rm e}^{i(p_0x_0+q_0y_0) + (p_i{\rm
e}^{\frac{q_0}{2\kappa}}x^i + q_i{\rm
e}^{-\frac{p_0}{2\kappa}}y^i)} \label{slawion1a} \\
&&~~~~~~~~~~~~~~~~~~~~~~~~~~~~~~~~~~\cdot\;\delta ( C^2_{\kappa}
(\vec{p}{\rm e}^{\frac{q_0}{2\kappa}},p_0) - M^2)~ \delta (
C^2_{\kappa} (\vec{q}{\rm e}^{-\frac{p_0}{2\kappa}},q_0) - M^2
)\,,\nonumber
\end{eqnarray}

ii) {\it $\kappa$-deformation of field oscillators algebra}. We see
from (\ref{slawion1a}) that the field oscillators in the product
${\hat \varphi}({x}){\hat \star}_{\kappa} {\hat \varphi}(y)$ are put
on modified mass-shells (see (\ref{slawion})). The unconventional
feature of such an approach is the use of quantized field
oscillators ${\hat A}(p_0,\vec{p})$ extended to the values of
$p=(p_0,\vec{p})$ which do not satisfy the $\kappa$-deformed
mass-shell condition (\ref{casimir})
%Interesting enough, in order to obtain the commutator (\ref{pauli})
%described by the momentum integral over $\kappa$-deformed mass-shell
%delta, it is necessary to employ the off-shell $\kappa$-deformed
%oscillators
\begin{equation}
{\hat A}(p_0,\vec{p})|_{ C^2_{\kappa} (\vec{p},p_0) = M^2}
\Rightarrow {\hat A}(p_0,\vec{p})|_{ C^2_{\kappa} (\vec{p},p_0) \ne
M^2}\label{frontln}\;.
\end{equation}
The pair of modified mass-shell conditions following from
(\ref{slawion}) is
\begin{equation}
C^2_{\kappa} (\vec{p}{\rm e}^{\frac{q_0}{2\kappa}},p_0) - M^2 =
0\;\;\;,\;\;\;C^2_{\kappa} (\vec{q}{\rm
e}^{-\frac{p_0}{2\kappa}},q_0) - M^2 =0 \;.\label{kkkkkkkkkkkk}
\end{equation}
Such a modification can be interpreted by the coproduct relation
(\ref{c4}), where $(p_0\,,\vec{p}{\rm e}^{q_0/2\kappa})$ and
$(q_0\,,\vec{q}{\rm e}^{-p_0/2\kappa})$ correspond to first and
second term in  the fourmomentum addition formula for the
two-particle state
\begin{equation}
P_0\triangleright
\hat{A}(p)\hat{A}(q)=(p_0+q_0)\hat{A}(p)\hat{A}(q)\,,
\end{equation}
\begin{equation}
P_i\triangleright \hat{A}(p)\hat{A}(q):=m\circ [\Delta(P_i)
\hat{A}(p)\otimes \hat{A}(q)]=[p_i {\rm e}^{q_0/2\kappa}+q_i {\rm
e}^{-p_0/2\kappa}]\hat{A}(p)\hat{A}(q)\,.
\end{equation}
where $m\circ(f\otimes g)=fg$. The energy values $p_0,q_0$
satisfying (\ref{kkkkkkkkkkkk}) are  equivalently described as the
solutions $p_0^{(\epsilon,\epsilon')}(\vec{p},\vec{q})$,
$q_0^{(\epsilon,\epsilon')}(\vec{p},\vec{q})$ ($\epsilon =\pm 1$,
$\epsilon'=\pm 1)$ of the following two coupled equations
\begin{equation}
p_0^{(\epsilon,\epsilon')}=\epsilon\omega_\kappa(\vec{p}{\rm
e}^{q_0^{(\epsilon,\epsilon')}/2\kappa})\;\;\;,\;\;\;
q_0^{(\epsilon,\epsilon')}=\epsilon'\omega_\kappa(\vec{q}{\rm
e}^{-p_0^{(\epsilon,\epsilon')}/2\kappa})\,. \label{ggg}
\end{equation}
We see from (\ref{ggg}) that for $\kappa$-deformed two-particle
states the energy of the first particle depends also on the
three-momenta of the energy, i.e. the modification (\ref{slawion})
of the mass-shell condition couples both constituents of the
two-particle state.

For particular choice of $\kappa$-deformed oscillator algebra  we
did show in \cite{1}, \cite{10} that using the modified
$\kappa$-star product (\ref{slawion1a}) one gets the $c$-number
field commutator.  One obtains
\begin{equation}
[\;\hat\varphi (x),\hat\varphi (y)\;]_{{\hat \star}_{\kappa}} =
{i}\Delta_\kappa (x-y;M^2)\;, \label{micro1}
\end{equation}
where
\begin{equation}
\Delta_\kappa (x;M^2) = \frac{i}{(2\pi)^3} \int d^4p \epsilon(p_0)
\delta \left(\left (2\kappa\sinh
\left({p_0}/{2\kappa}\right)\right)^2 - \vec{p}^{\ 2} - M^2 \right)
{\rm e}^{i{p}_\mu{x}^{\mu}}\;, \label{pauli}
\end{equation}
is the $\kappa$-deformed Pauli-Jordan commutator function proposed
firstly in \cite{9} after using somewhat naive arguments.

b) {\it Plan of the paper and the resume of results}. In our earlier
papers \cite{11}, \cite{1}, \cite{10} we studied a definite model of
$\kappa$-deformed oscillators, without any free parameter (except
the deformation parameter $\kappa$). In this paper we consider the
most general set of binary $\kappa$-deformed oscillator algebras
consistent with $\kappa$-deformed addition law of the fourmomenta.

 Firstly, in Sect. 2 we
shall use  $\kappa$-star product (\ref{newstar1}) and the
oscillators ${\hat A}(p_0,\vec{p})$ which lie  on standard
$\kappa$-deformed mass-shells. We shall show  that in such a case
for any possible choice of binary oscillator algebra it is not
possible to obtain the $c$-number value of the field commutator. In
place of the formula (\ref{micro1})
 one gets the $q$-number field commutator  bilinear in the
$\kappa$-deformed oscillators. In particular we shall discuss
briefly two recent proposals of $\kappa$-deformed oscillator
algebras (\cite{12}, \cite{13}) which fall into such a category.

In Sect. 3 we shall consider the binary  multiplication of
$\kappa$-deformed free fields using the modified  $\kappa$-deformed
$\star$-product. In such a case by performing the general
transformation in two-particle fourmomentum space we arrive at large
class of $\kappa$-deformed oscillator algebras depending on six
arbitrary functions of two-particle fourmomenta, which all  lead to
the $c$-number field commutator. All such $\kappa$-oscillators are
characterized by modified energy-momentum dispersion relation
generalizing the relations (\ref{kkkkkkkkkkkk}) or (\ref{ggg}) and
can be classified  by various  forms of the addition law for the
threemomenta of $\kappa$-deformed two-particle state. The particular
choice proposed in \cite{11}, \cite{1} was  described by the Abelian
addition law
\begin{equation}
\vec{p}_{1+2} = \vec{p}+ \vec{q}  \;,\label{nowywzor}
\end{equation}
while the  choices from \cite{12}, \cite{13} (see also \cite{go})
were  characterized by the non-Abelian addition  formula\footnote{We
recall that we use the $\kappa$-Poincare algebra which leads to the
standard (see \cite{8}) coproduct (\ref{c4}) for the threemomenta.}
\begin{equation}
\vec{p}_{1+2} = \vec{p}\dotplus \vec{q} = \vec{p} {\rm
e}^{\frac{q_0}{2\kappa}} + \vec{q}{\rm
e}^{-\frac{p_0}{2\kappa}}\;.\label{csa4}
\end{equation}
The $\kappa$-deformed oscillator algebra with the threemomentum
addition law (\ref{csa4}) was constructed by the use of suitable
deformation of the flip operator. In Sect. 4 we describe the general
oscillator algebras (\ref{aallgge}) (see Sect. 3) as  the
composition of the most general $\kappa$-deformed multiplication and
the $\kappa$-deformed flip operation.

The $\kappa$-deformed field oscillators determine the structure of
corresponding $\kappa$-deformed multiparticle states\footnote{The
discussion of multiparticle states for $n>2$ has been recently
considered in \cite{ost}.}. We shall only mention here that the
n-particle sector of $\kappa$-deformed Fock space is represented by
suitably constructed non-factorizable clusters. We interpret such
structure of $\kappa$-deformed Fock spaces as a result of  the
geometric interactions implied by the Lie-algebraic noncommutativity
of $\kappa$-deformed space-time. It would be interesting to
understand  how such
 features can be linked with quantum gravity framework.

\section{Standard $\kappa$-star product and the $\kappa$-deformed quantum fields}

The aim of this section is to study the commutator of the fields
(\ref{field}) with the $\kappa$-star multiplication rule
(\ref{newstar1})
\begin{eqnarray}
&&[\,{\hat \varphi}({x}), {\hat
\varphi}({y})\,]_{\star_{\kappa}}=\label{zadrugizm111}\\
&&\qquad=\frac{1}{(2\pi)^{3}}\int d^4p\,d^4q\, {\hat A}(p){\hat
A}(q){\rm e}^{ipx}\star_{\kappa}
{\rm e}^{iqy}\,\delta(C^2_{\kappa}(p)-M^2)\delta(C^2_{\kappa}(q)-M^2)\nonumber\\
&&\qquad-\frac{1}{(2\pi)^{3}}\int d^4p'\,d^4q'\,{\hat A}(q'){\hat
A}(p'){\rm e}^{iq'y}\star_{\kappa} {\rm
e}^{ip'x}\,\delta(C^2_{\kappa}(p')-M^2)\delta(C^2_{\kappa}(q')-M^2)\;.\nonumber
\end{eqnarray}
We see from (\ref{zadrugizm111}) that due to the presence of
respective Dirac deltas the field oscillators remain on
$\kappa$-deformed mass-shell (\ref{casimir}). We shall look for the
generalized binary relations of $\kappa$-deformed oscillators
consistent with the $\kappa$-deformed fourmomentum addition law. We
shall show that for any choice of these binary relations determining
the choice of $\kappa$-statistics, the commutator
(\ref{zadrugizm111}) is an operator bilinear in the field
oscillators. We use in this paper the definition of (generalized)
free field as characterized by the $c$-number commutator function
(see e.g. \cite{cnumber1}, \cite{cnumber2}). We obtain therefore in
this section the result that using standard on-shell
$\kappa$-oscillators ${\hat A}(p_0,\vec{p})$ we can  not obtain the
free $\kappa$-deformed quantum fields.

Our demonstration  of the operator nature of the commutator
(\ref{zadrugizm111}) follows from the impossibility of the
factorization under momenta integrals of any  binary relations for
the field oscillators ${\hat A}(p_0,\vec{p})$. For studying possible
factorization we
shall perform  the following general O(3)-covariant  change of variables separately  %$(p_0 =
%\mathcal{P}_0$, $q_0 = \mathcal{Q}_0)$
in the first term on rhs of (\ref{zadrugizm111})
$(\tilde{f}=\tilde{f}(p,q)$ etc)
\begin{equation}
  \vec{p}\rightarrow\vec{\mathcal{P}}(p,q)=\vec{p}\tilde{f}+\vec{q}\tilde{g}\;\;\;,\;\;\;
  \vec{q}\rightarrow\vec{\mathcal{Q}}(p,q)=\vec{p}\tilde{k}+\vec{q}\tilde{l}\;,\label{change}
 \end{equation}
\begin{equation}
 {p}_0\rightarrow{\mathcal{P}}_0(p,q)\;\;\;,\;\;\;
  {q}_0\rightarrow{\mathcal{Q}}_0(p,q)\;,\label{newchange}
 \end{equation}
 with the following inverse formulae $(f=f(\mathcal{P},\mathcal{Q})$ etc)
 \begin{equation}
  \vec{\p}\rightarrow\vec{p}(\p,\q)=\vec{\mathcal{P}}f+\vec{\mathcal{Q}}g\;\;\;,\;\;\;
  \vec{\q}\rightarrow\vec{q}(\p,\q)=\vec{\mathcal{P}}k+\vec{\mathcal{Q}}l\,,\label{changeinv1}
  \end{equation}
  \begin{equation}
  \p_0\rightarrow p_0(\p,\q)\,\;\;\;,\;\;\;\q_0\rightarrow q_0(\p,\q)\,, \label{changeinv1a}
 \end{equation}
and in the second term on rhs of (\ref{zadrugizm111})
\begin{equation}
\vec{p'}\rightarrow\vec{\mathcal{P}}'=\vec{p'}\tilde{f}'+\vec{q'}\tilde{g}'\;\;\;,\;\;\;
\vec{q'}\rightarrow\vec{\mathcal{Q}}'=\vec{p'}\tilde{k}'+\vec{q'}\tilde{l}'\;,
\label{change1}
\end{equation}
\begin{equation}
{p}_0'\rightarrow{\mathcal{P}}'_0\;\;\;,\;\;\;
{q}_0'\rightarrow{\mathcal{Q}}'_0\;, \label{newchange1}
\end{equation}
with the inverse formulae
\begin{equation}
  \vec{\p}'\rightarrow\vec{p'}(\p',\q')=\vec{\mathcal{P}}'f'+\vec{\mathcal{Q}}'g'\;\;\;,\;\;\;
  \vec{\q}'\rightarrow\vec{q'}(\p',\q')=\vec{\mathcal{P}}'k'+\vec{\mathcal{Q}}'l'\,, \label{changeinv2}
  \end{equation}
  \begin{equation}
  \p_0\rightarrow p_0'(\p',\q')\,\;\;\;,\;\;\;\q_0\rightarrow q_0'(\p',\q')\,. \label{changeinv2a}
 \end{equation}
We assume that $f, g, h, k, p_0,q_0$ in (\ref{changeinv1}) and $ f',
g', h', k', p_0',q_0'$ in (\ref{changeinv2})  are respectively
arbitrary O(3)-invariant functions of the fourmomenta
$\p=(\vec{\p},\p_0)$, $\q=(\vec{\q},\q_0)$ and
$\p'=(\vec{\p}',\p_0')$, $\q'=(\vec{\q'},\q_0')$\footnote{We recall
that the classical SO(3) Hopf algebra is a sub-Hopf algebra of
$\kappa$-deformed Poincare Hopf algebra.}.

We shall look for such a choice of arbitrary functions in the
formulae (\ref{changeinv1}),(\ref{changeinv1a}) and
(\ref{changeinv2}), (\ref{changeinv2a}) which leads to the equality
%(${\mathcal{P}x} \equiv \vec{\mathcal{P}}x - \mathcal{P}_0x_0$ etc.)
\begin{eqnarray}
{\rm e}^{ip(\p,\q)x}\star_{\kappa}{\rm e}^{iq(\p,\q)y} = {\rm
e}^{iq'(\p,\q)y}\star_{\kappa}{\rm e}^{ip'(\p,\q)x} \label{equality}
\;,
\end{eqnarray}
with $\kappa$-deformed mass-shell conditions taken into account. The
formula (\ref{equality}) describes a necessary condition which
permits to factorize in the commutator (\ref{zadrugizm111}) the
oscillator algebra and to derive the $c$-number commutator function.
We should observe that the change of variables
(\ref{change})-(\ref{changeinv2}) modifies as well the explicit form
of the product of mass-shell deltas in (\ref{zadrugizm111}). We
obtain in first term of rhs of (\ref{zadrugizm111})
\begin{eqnarray}
&&\delta(C^2_{\kappa} (p) - M^2)\cdot \delta( C^2_{\kappa} (q) -
M^2) \to \cr &&~~~~~~~~~~~~~~~~\to \delta( C^2_{\kappa} (p(\p,\q)) -
M^2)\cdot \delta(C^2_{\kappa} (q(\p,\q)) - M^2) \label{deltas1} \;,
\end{eqnarray}
and in the second term we get
\begin{eqnarray}
&&\delta( C^2_{\kappa} (p') - M^2)\cdot \delta( C^2_{\kappa} (q') -
M^2) \to \cr &&~~~~~~~~~~~~~~~~\to \delta( C^2_{\kappa} (p'(\p,\q))
- M^2)\cdot \delta( C^2_{\kappa} (q'(\p,\q)) - M^2) \label{deltas2}
\;.
\end{eqnarray}

One obtains\footnote{Because in transformed formula
(\ref{zadrugizm111}) we integrate over the variables $\p',\q'$,
further we shall denote them similarly as in first term of the
commutator by $\p$ and $\q$.}
\begin{eqnarray}
&&[\,{\hat \varphi}({x}), {\hat
\varphi}({y})\,]_{\star_{\kappa}}=\frac{1}{(2\pi)^{3}}\int
d^4\p\,d^4\q\, J\left(\stackrel{p,\ q}{\p, \q}\right)
 {\hat A}(p(\p,\q)){\hat A}(q(\p,\q))\label{commutd}\\
&&~~~~~~~~~~~~~~~~~~~~~~~~~~~~~~~~~~~~~~~~\cdot {\rm e}^{ip(\p,\q)x}\star_{\kappa}{\rm e}^{iq(\p,\q)y}\nonumber\\
&&~~~~~~~~~~~~~~~~~~~~~~~~~~~~~~~~~~~~~~~~\cdot \delta( C^2_{\kappa}
(p(\p,\q)) - M^2)\cdot
\delta(C^2_{\kappa} (q(\p,\q)) - M^2)\nonumber\\
&&~~~~~~~~~~~~~~~~~~-\frac{1}{(2\pi)^{3}}\int
d^4\p\,d^4\q\,J\left(\stackrel{p',\ q'}{\p,\q}\right)
 {\hat A}(q'(\p,\q)){\hat A}(p'(\p,\q))\nonumber\\
&&~~~~~~~~~~~~~~~~~~~~~~~~~~~~~~~~~~~~~~~~\cdot {\rm
e}^{iq'(\p,\q)y}\star_{\kappa}
{\rm e}^{ip'(\p,\q)x}\nonumber\\
&&~~~~~~~~~~~~~~~~~~~~~~~~~~~~~~~~~~~~~~~~\cdot \delta( C^2_{\kappa}
(p'(\p,\q)) - M^2)\cdot \delta(C^2_{\kappa} (q'(\p,\q)) -
M^2)\,.\nonumber
\end{eqnarray}

Let us denote by $\p_0 = \pi(\vec{\p},\vec{\q})$, $\q_0 =
\rho(\vec{\p},\vec{\q})$ and $\p_0' = \pi'(\vec{\p},\vec{\q})$,
$\q_0' = \rho'(\vec{\p},\vec{\q})$ the solutions of the following
two pairs of the deformed mass-shell conditions
\begin{eqnarray}
 C^2_{\kappa}(p(\p,\q))
 - M^2=0\;\;\;,\;\;\; C^2_{\kappa}(q(\p,\q) )
 - M^2=0\label{cond1} \;,
\end{eqnarray}
\begin{eqnarray}
 C^2_{\kappa}(p'(\p,\q))
 - M^2=0\;\;\;,\;\;\; C^2_{\kappa}(q'(\p,\q) )
 - M^2=0\label{cond2} \;.
\end{eqnarray}
%corresponding to two terms in the commutator (\ref{zadrugizm111}).
%In fact, in  relation (\ref{equality}) which is multiplied side wise
%by mass-shell deltas with the arguments (\ref{cond1}),
%(\ref{cond2}), one should substitute  the values
%\begin{eqnarray}
%\mathcal{P}_0 = \mathcal{P}_0^{L}\;\;\;,\;\;\;
%\mathcal{Q}_0=\mathcal{Q}_0^{L}\;\;\;,\;\;\;\tilde{\mathcal{P}}_0 =
%\mathcal{P}_0^{R}\;\;\;,\;\;\;\tilde{\mathcal{Q}}_0 =
%\mathcal{Q}_0^{R}\;.\label{wspolnotatworzaca}
%\end{eqnarray}
%
We shall consider firstly the validity of restricted relation
(\ref{equality})   obtained by putting  $x_0 = y_0 = 0$. It is easy
to check that one gets the equality of deformed threemomenta
exponentials
\begin{eqnarray}
&&{\rm e}^{ip(\p,\q)x}\star_{\kappa}{\rm e}^{iq(\p,\q)y}|_{\p_0=\pi,\q_0=\rho,\,(x_0=y_0=0)}\label{eqqqau}\\
&&~~~~~~~~~~~~~~~~~~~~~~~~=\exp\left[{i(\vec{p}(\p,\q){\rm
e}^{\frac{q_0(\p,\q)}{2\kappa}}\vec{x} + \vec{q}(\p,\q)}{\rm
e}^{-\frac{p_0(\p,\q)}{2\kappa}}\vec{y}) \right]|_{\p_0=\pi,\q_0=\rho}\nonumber\\
&&~~~~~~~~~~~~~~~~~~~~~~~~=\exp\left[{i(\vec{q'}(\p,\q){\rm
e}^{\frac{p_0'(\p,\q)}{2\kappa}}\vec{y} + \vec{p'}(\p,\q)}{\rm
e}^{-\frac{q_0'(\p,\q)}{2\kappa}}\vec{x}) \right]|_{\p_0=\pi',\q_0=\rho'}\nonumber\\
&&={\rm e}^{iq'(\p,\q)y}\star_{\kappa}{\rm
e}^{ip'(\p,\q)x}|_{\p_0'=\pi',\q_0'=\rho',\,(x_0=y_0=0)}\nonumber\;,
\end{eqnarray}
if the arbitrary functions introduced in the relations
(\ref{changeinv1}),(\ref{changeinv1a}) and
(\ref{changeinv2}),(\ref{changeinv2a}) satisfy the relations
 \begin{eqnarray}
&&f(\p,\q){\rm e}^{p_0(\p,\q)/2\kappa}=f'(\p,\q){\rm
e}^{-p_0'(\p,\q)/2\kappa}\;\;,\;\;g(\p,\q){\rm
e}^{p_0(\p,\q)/2\kappa}=
 g'(\p,\q){\rm e}^{-p_0'(\p,\q)/2\kappa}\;,\nonumber\\
&& k(\p,\q){\rm e}^{-q_0(\p,\q)/2\kappa}=
 k'(\p,\q){\rm e}^{q_0'(\p,\q)/2\kappa}\;\;,\;\; l(\p,\q){\rm e}^{-q_0(\p,\q)/2\kappa}=l'(\p,\q){\rm e}^{q_0'(\p,\q)/2\kappa}\;,\qquad
 \label{12}
 \end{eqnarray}
and $\p_0=\pi$, $\q_0=\rho$, $\p_0=\pi'$,
 $\q_0=\rho'$ satisfy the mass-shell conditions
 (\ref{cond1}),(\ref{cond2}).

 If we assume the  relations (\ref{12}), then the formula (\ref{equality})
reduces to  the equality of  time exponentials
\begin{equation}
{\rm e}^{i[x_0p_0(\p,\q) + y_0q_0(\p,\q)]}|_{\p_0=\pi,\q_0=\rho}
={\rm e}^{i[x_0p_0'(\p,\q) +
y_0q_0'(\p,\q)]}|_{\p_0=\pi',\q_0=\rho'}\;.\label{dadzbog}
\end{equation}
The relation (\ref{dadzbog}) is satisfied for any values of $x_0$,
$y_0$  only if
\begin{equation}
p_0(\vec{\p},\pi;\vec{\q},\rho)  =
p_0'(\vec{\p},\pi';\vec{\q},\rho')\;\;\;,\;\;\;
q_0(\vec{\p},\pi;\vec{\q},\rho)  =
q_0'(\vec{\p},\pi';\vec{\q},\rho')\;.\label{dadzbog1}
\end{equation}
In order to derive the restrictions following from (\ref{dadzbog1})
one can rewrite the relations (\ref{cond1}),(\ref{cond2}) in the
form of the following identities ($\epsilon = \pm 1$ etc)
\begin{equation}
p_0(\vec{\p},\pi;\vec{\q},\rho)=\epsilon\;\omega_\kappa(\vec{p}^{\
2}(\vec{\p},\pi;\vec{\q},\rho))\;\;\;,\;\;\;
q_0(\vec{\p},\pi;\vec{\q},\rho)=\eta\;\omega_\kappa(\vec{q}^{\
2}(\vec{\p},\pi;\vec{\q},\rho))\,, \label{ds1}
\end{equation}
\begin{equation}
p'_0(\vec{\p},\pi;\vec{\q},\rho)=\epsilon'\;\omega_\kappa(\vec{p'}^{\
2}(\vec{\p},\pi;\vec{\q},\rho))\;\;\;,\;\;\;
q'_0(\vec{\p},\pi;\vec{\q},\rho)=\eta'\;\omega_\kappa(\vec{q'}^{\
2}(\vec{\p},\pi;\vec{\q},\rho))\,, \label{ds2}
\end{equation}
where $\omega_\kappa(\vec{p})$ is defined by relation (\ref{omega}).
We see from (\ref{ds1}),(\ref{ds2}) that the relation
(\ref{dadzbog1}) can be written as follows\footnote{Because
$\omega_\kappa\geq0$ the relations (\ref{dadzbog1}) require that
$\epsilon=\epsilon'$ and $\eta=\eta'$.}
\begin{equation}
\omega_\kappa(\vec{p}^{\
2}(\vec{\p},\pi;\vec{\q},\rho))=\omega_\kappa(\vec{p'}^{\
2}(\vec{\p},\pi;\vec{\q},\rho))\,, \label{gf1}
\end{equation}
\begin{equation}
\omega_\kappa(\vec{q}^{\
2}(\vec{\p},\pi;\vec{\q},\rho))=\omega_\kappa(\vec{q'}^{\
2}(\vec{\p},\pi;\vec{\q},\rho))\,, \label{gf2}
\end{equation}
which implies  that
\begin{equation}
\vec{p}^{\ 2}(\vec{\p},\pi;\vec{\q},\rho)=\vec{p'}^{\
2}(\vec{\p},\pi;\vec{\q},\rho)\;\;\;,\;\;\; \vec{q}^{\
2}(\vec{\p},\pi;\vec{\q},\rho)=\vec{q'}^{\
2}(\vec{\p},\pi;\vec{\q},\rho)\,.\label{najjjka}
\end{equation}
Inserting in first formula of (\ref{najjjka}) the relations
(\ref{changeinv1}),(\ref{changeinv1a}) and
(\ref{changeinv2}),(\ref{changeinv2a}) one gets from (\ref{gf1}) the
condition
\begin{equation}
f^2\vec{\p}^2+2fg\vec{\p}\vec{\q}+g^2\vec{\q}^2=f'^2\vec{\p}^2+2f'g'\vec{\p}\vec{\q}+g'^2\vec{\q}^2\,,
\label{bbb}
\end{equation}
and analogous relation obtained by the replacements $\vec{p} \to
\vec{p}'$ and $\vec{q} \to \vec{q}'$ in (\ref{gf2}). The insertion
of (\ref{12}) into (\ref{bbb}) leads to
\begin{equation}
(1-{\rm e}^{(p_0+p'_0)/2\kappa})\vec{p}^{\
2}(\vec{\p},\pi;\vec{\q},\rho)=0\,.
\end{equation}
Because $\vec{p}^{\ 2}$ is positive-definite, the condition
(\ref{bbb})
 requires for finite $\kappa$ that $p_0(\vec{\p},\pi;\vec{\q},\rho)  = -p_0'(\vec{\p},\pi';\vec{\q},\rho')\,,
q_0(\vec{\p},\pi;\vec{\q},\rho)  =
-q_0'(\vec{\p},\pi';\vec{\q},\rho')$ what contradicts the relations
(\ref{dadzbog1}).

In conclusion, both relations (\ref{12}) and (\ref{dadzbog1})
required for the validity of the relation (\ref{equality}) can not
be valid, and consequently if $x_0 \ne y_0$ for on-shell
oscillators, it is not possible to factorize in the
$\kappa$-deformed commutator function (\ref{zadrugizm111}) any
$\kappa$-deformed oscillator algebra.
\newline

 {\it Examples in recent literature.} In recent papers \cite{12}-\cite{ny} the following particular
choice has been made
\begin{eqnarray}
\vec{p} =\vec{\p}\;\;\;(f=1,\; g=0)\;\;\;\;,\;\;\;\;\vec{q}
=\vec{\q}\;\;\;\;\;(l=1,\;
k=0)\;\;\;\;,\;\;\;\;p_0=\p_0\;\;\;\;,\;\;\;\; q_0=\q_0\,.
\label{choice}
\end{eqnarray}
 The restrictions
on functions $\vec{p'}(\p,\q)$, $\vec{q'}(\p,\q)$ were obtained from
non-Abelian composition law of threemomenta (see (\ref{csa4})),
considered as an identity in the variables $\vec{\p}$, $\vec{\q}$
\begin{equation}
 \vec{\p} \dotplus \vec{\q} = \vec{q'}(\p,\q)
 \dotplus \vec{p'}(\p,\q)\;,\label{jadroslawi}
\end{equation}
where $\dotplus$ denotes addition law based on $\kappa$-deformed
fourmomentum coproduct. Second relation follows  from the energy conservation law. \\
For our choice of the threemomenta coproduct (see (\ref{c4})) eq.
(\ref{jadroslawi}) takes the following explicit form
\begin{equation}
 \vec{\p}{\rm
e}^{\frac{\p_0}{2\kappa}} +\vec{\q}{\rm e}^{-\frac{\q_0}{2\kappa}}
 = \vec{q'}(\p,\q){\rm
e}^{\frac{{p'_0}}{2\kappa}} +\vec{p'}(\p,\q){\rm
e}^{-\frac{{q'_0}}{2\kappa}} \;,\label{jadroslawi1}
\end{equation}
where $\p_0 = \pm \omega_{\kappa}(\vec{\p})$, $\q_0 = \pm
\omega_{\kappa}(\vec{\q})$ and the relations (\ref{ds2}) should be
inserted on rhs of (\ref{jadroslawi1}). One gets the relation
(\ref{jadroslawi1}) as identity in particular if\footnote{In
\cite{12}-\cite{ny} there is used different bicrossproduct basis of
$\kappa$-deformed Poincare algebra, what leads to the coproduct
$\Delta(\vec{P}') = \vec{P}'\otimes 1 + {\rm
e}^{-\frac{P_0}{\kappa}}\otimes \vec{P}'$ and suitable modification
of our formulae (\ref{jadroslawi1}),(\ref{swarzyca400e}).}
\begin{equation}
 \vec{p'}(\p,\q) = \vec{\p} {\rm exp}\left(\frac{\p_0 + q'_0}{2\kappa}\right)
\;\;\;,\;\;\;\vec{q'}(\p,\q) = \vec{\q} {\rm exp}\left(-\frac{\q_0 +
p'_0}{2\kappa}\right) \;,\label{swarzyca400e}
\end{equation}
with the relations (\ref{ds2}) taking the following explicite form
($\epsilon'=\pm 1\,, \eta'=\pm 1$)
\begin{equation}
 p'_0 = \epsilon' \omega_{\kappa}
 (\vec{\p}{\rm e}^{\frac{\p_0 + q'_0}{2\kappa}})
\;\;\;,\;\;\; q'_0 = \eta' \omega_{\kappa}
 (\vec{\q}{\rm e}^{-\frac{(\q_0 + p'_0)}{2\kappa}})
\;,\label{swarzyca400f}
\end{equation}
determining $p'_0 = p'_0(\vec{\p},\vec{\q})$ and $q'_0 =
q'_0(\vec{\p},\vec{\q})$ as functions of threemomenta $\vec{\p}$ and
$\vec{\q}$. The energy conservation relation  takes the form
\begin{equation}
\p_0+\q_0 = p'_0 + q'_0 \Leftrightarrow
\epsilon\omega_{\kappa}(\vec{\p}) + \eta\omega_{\kappa}(\vec{\q})
=\epsilon' \omega_{\kappa}
 (\vec{\p}{\rm e}^{\frac{\p_0 + q'_0}{2\kappa}})+
 \eta' \omega_{\kappa}
 (\vec{\q}{\rm e}^{-\frac{(\q_0 + p'_0)}{2\kappa}})\,.\label{swarzyca400g}
\end{equation}
It should be stressed however that even if the relations
(\ref{swarzyca400g}) are valid the relations (\ref{dadzbog1})
crucial for obtaining the $c$-number commutator can not satisfied.

One can point out that  equations  (\ref{jadroslawi1}) do not
specify completely the six functions $p'_i(\p,\q)$, $q'_i(\p,\q)$.
In \cite{13}, \cite{ny} it has been additionally assumed that two
products of the oscillators ${\hat A}(\vec{\p}){\hat A}(\vec{\q})$
and multiplied in flipped order ${\hat A}(\vec{q'}\,,q_0'){\hat
A}(\vec{p'}\,,p_0')$ transform in the same covariant way under the
$\kappa$-deformed boost generators. It was shown that:

i) for $D=2$ $\kappa$-deformed system there exist a unique
$\kappa$-covariant choice of functions $\vec{p'}$, $\vec{q'}$
consistent with relations (\ref{swarzyca400g}) and $D=2$ counterpart
of (\ref{jadroslawi1}),

ii) in $D=4$ under analogous assumptions the solution is known only
in lowest three orders of $\frac{1}{\kappa}$ perturbation expansion.

The results obtained in \cite{13}, \cite{ny} providing the
$\kappa$-covariant set of $\kappa$-statistics   are very
interesting, but because they describe the states with
 the fourmomenta satisfying standard on-shell conditions
(\ref{cond1}),(\ref{cond2}) they are not consistent with  the
relation (\ref{dadzbog1}) which is necessary for obtaining the field
commutator as a $c$-number.

\section{Modified  $\kappa$-deformed star product and free
$\kappa$-deformed quantum fields}

We see from Sect. 2  that the main problem in obtaining $c$-number
field commutators is the difficulty with  getting valid the
relations (\ref{dadzbog1}). We recall  that in Sect. 2 we obtained
different on-shell values of $p_0(\p,\q)$, $q_0(\p,\q)$ and
$p'_0(\p,\q)$, $q'_0(\p,\q)$ following  from different forms of the
$\kappa$-deformed mass-shell conditions (\ref{cond1}) and
(\ref{cond2}). In this Section we shall change  the
$\kappa$-deformed star product (\ref{newstar1}) in such a way that
both  modified mass-shell conditions (\ref{cond1}) and (\ref{cond2})
will  become identical. We  will provide two pairs of  the same
modified energy-momentum dispersion relations for $p_0(\p,\q)$,
$p'_0(\p,\q)$ and $q_0(\p,\q)$, $q'_0(\p,\q)$, what allows  the
validity of the relations (\ref{dadzbog1}) and $c$-number
commutation function.

For that purpose we shall use  modified $\kappa$-deformed star
product (\ref{slawion1a}). The corresponding field commutator of
$\kappa$-deformed free fields (\ref{field}) takes the form
\begin{eqnarray}
 &&[\;{\hat \varphi}(x)\,,{\hat \varphi}(y)\;]_{{\hat \star}_\kappa}=
 \frac{1}{(2\pi)^{3}} \int d^4p\,d^4q\, {\hat A}(p){\hat A}(q){\rm e}^{ipx}
 \star_\kappa {\rm e}^{iqy}\label{zap}\\
&&\qquad\qquad\qquad\qquad\qquad\qquad\cdot\delta(C^2_\kappa(\vec{p}{\rm
e}^{q_0/2\kappa},p_0)-M^2)
 \delta(C^2(\vec{q}{\rm e}^{-p_0/2\kappa},q_0)-M^2)\nonumber\\
 &&\;\;\qquad\qquad\;\;\qquad-\frac{1}{(2\pi)^{3}}
 \int d^4p'\,d^4q'\,{\hat A}(q'){\hat A}(p'){\rm e}^{iq'y}\star_\kappa {\rm e}^{ip'x}\nonumber\\
 &&\qquad\qquad\qquad\qquad\qquad\qquad\cdot\delta(C^2_\kappa(q'_0,\vec{q'}{\rm e}^{p'_0/2\kappa})-M^2)
 \delta(C^2(p'_0,\vec{p'}{\rm e}^{-q'_0/2\kappa})-M^2)\;.\nonumber
 \end{eqnarray}
Introducing the change of momentum variables
(\ref{change})-(\ref{changeinv2a}) one obtains the formula
\begin{eqnarray}
&&[\;{\hat \varphi}(x)\,,{\hat \varphi}(y)\;]_{{\hat
\star}_\kappa}=\label{18}\\
&&\qquad\qquad=\frac{1}{(2\pi)^{3}}\int d^4\p\,d^4\q\,
J\left(\stackrel{p,\ q}{\p,\q}\right)
\hat{A}(p_0\,,\vec{\p}f+\vec{\q}g)\hat{A}(q_0\,,\vec{\p}k+\vec{\q}l)\,\cr
&&~~~~~~~~~~~~~\qquad\cdot\,\exp{[i(p_0x^0+q_0y^0)]}
\,\exp{[-i[(\vec{\p}f+\vec{\q}g){\rm
e}^{q_0/2\kappa}\vec{x}+(\vec{\p}k+\vec{\q}l){\rm
e}^{-p_0/2\kappa}\vec{y}]]} \, \cr
&&~~~~~~~~~~~~~\qquad\cdot\,\delta(C^2_\kappa(p_0\,,[\vec{\p}f+\vec{\q}g]{\rm
e}^{q_0/2\kappa})-M^2)
\delta(C^2_\kappa(q_0\,,[\vec{\p}k+\vec{\q}l]{\rm
e}^{-p_0/2\kappa})-M^2)
\nonumber\\
 &&~~~~~~~~~~~-\frac{1}{(2\pi)^{3}}\int d^4\p\,d^4\q\,
 J\left(\stackrel{p',\ q'}{\p,\q}\right)
 \hat{A}(q_0'\,,\vec{\p}k'+\vec{\q}l')\hat{A}(p_0'\,,\vec{\p}f'+\vec{\q}g')
 \cr
&&~~~~~~~\qquad\qquad\cdot\,\exp{i(p_0'x^0+q_0'y^0)}
\exp[{-i[(\vec{\p}k'+\vec{\q}l'){\rm e}^{p_0'/2\kappa}\vec{y}+
 (\vec{\p}f'+\vec{\q}g'){\rm e}^{-q_0'/2\kappa}\vec{x}]}]\,\nonumber\\
 &&~~~~~~~~~~~~~\qquad\cdot\,\delta(C^2_\kappa(q'_0\,,[\vec{\p}k'+\vec{\q}l'] {\rm e}^{p_0'/2\kappa})-M^2)
 \,\delta(C^2(p_0'\,,[\vec{\p}f'+\vec{\q}g']{\rm e}^{-q'_0/2\kappa})-M^2)\;,\nonumber
\end{eqnarray}
where $J\left(\stackrel{p,\ q}{\p,\q}\right)$,
$J\left(\stackrel{p',\ q'}{\p,\q}\right)$ describe respectively the
Jacobians of transformations (\ref{changeinv1}),(\ref{changeinv1a})
and (\ref{changeinv2}),(\ref{changeinv2a}). We see that by
replacement (\ref{slawion}) we matched in two consecutive terms in
(\ref{zap}) the asymmetry of the star product of exponentials with
the asymmetry of   mass-shell deltas. After the substitution of
relations (\ref{12})  (expressing $f',\;g',\;k',\;l',\;p_0',\;q_0'$
by ${f},\;{g},\;{k},\;{l},\;p_0,\;q_0$) we see that in (\ref{18})
the products of two  mass-shell deltas in two consecutive terms are
becoming  the same, and one can proceed to factorize the binary
algebraic relations describing the $\kappa$-deformed oscillator
algebra. After inserting the relations (\ref{12}) the energy values
$p_0$, $p_0'$ and $q_0$, $q_0'$ will satisfy the same mass-shell
conditions and therefore it will be  consistent to assume the
relations
(\ref{dadzbog1}). %We add also that from  the relations
%(\ref{dadzbog1}) follows the Abelian addition law (\ref{pedy})
%satisfied as identity.

The relations (\ref{dadzbog1}) provide  necessary conditions for the
factorization of $\kappa$-deformed algebra. One gets
\begin{eqnarray}
&&[\;{\hat \varphi}(x)\,,{\hat \varphi}(y)\;]_{{\hat
\star}_\kappa}=\int d^4\p\,d^4\q\, \left[ J\left(\stackrel{p,\
q}{\p,\q}\right) \hat{A}(p_0\,,\vec{\p}f+\vec{\q}g){\hat
A}(q_0\,,\vec{\p}k+\vec{\q}l)+\right.\label{19}\\
 &&~~~~~~~~~~~~~~~~~~~~~~~~~~~~\qquad
\left. - J\left(\stackrel{p',\ q'}{\p,\q}\right)
\hat{A}(q_0\,,(\vec{\p}{k}+\vec{\q}{l}){\rm e}^{-p_0/\kappa}){\hat
A}(p_0\,,(\vec{\p}{f}+\vec{\q}{g}){\rm
e}^{q_0/\kappa})\right]\nonumber
 \\
&&~~~~~~~~~~~~\qquad\cdot\,\exp{[i(p_0x^0+q_0y^0)]}\exp({-i[(\vec{\p}f+\vec{\q}g){\rm
e}^{q_0/2\kappa}
\vec{x}+(\vec{\p}k+\vec{\q}l){\rm e}^{-p_0/2\kappa}\vec{y}]})\,\nonumber\\
&&~~~~~~~~~~~~\qquad\cdot\delta(C^2_\kappa(p_0\,,[\vec{\p}f+\vec{\q}g]{\rm
e}^{q_0/2\kappa})-M^2)\,\delta(C^2_{\kappa}(q_0\,,[\vec{\p}k+\vec{\q}l]{\rm
e}^{-p_0/2\kappa})-M^2)\;.\nonumber
\end{eqnarray}
Under the integral (\ref{19})
$\p_0\equiv\tilde{\pi}(\vec{\p},\vec{\q})$ and
$\q_0\equiv\tilde{\rho}(\vec{\p}\,\vec{\q})$ describe respectively
 the solutions of the following coupled pair of modified
$\kappa$-deformed mass-shell conditions
\begin{eqnarray}
C^2_\kappa(p_0\,,[\vec{\p}f+\vec{\q}g]{\rm e}^{q_0/2\kappa})-M^2 =
0\;\;\;,\;\;\;C^2_{\kappa}(q_0\,,[\vec{\p}k+\vec{\q}l]{\rm
e}^{-p_0/2\kappa})-M^2 = 0\;. \label{equatttion}
\end{eqnarray}

In order to obtain the $c$-number value of the commutator (\ref{18})
we should postulate the following general $\kappa$-deformed
oscillator algebra
\begin{eqnarray}
&& J\left(\stackrel{p,\ q}{\p,\q}\right)
\hat{A}(p_0\,,\vec{\p}f+\vec{\q}g){\hat
A}(q_0\,,\vec{\p}k+\vec{\q}l)\label{aallgge}\\
 &&~~\qquad
 -
J\left(\stackrel{p',\ q'}{\p,\q}\right)
\hat{A}(q_0\,,(\vec{\p}{k}+\vec{\q}{l}){\rm e}^{-p_0/\kappa}){\hat
A}(p_0\,,(\vec{\p}{f}+\vec{\q}{g}){\rm e}^{q_0/\kappa})=c-{\rm
number}\,,\nonumber
\end{eqnarray}
where the functions
$p_0=p_0(\tilde{\pi},\vec{\p};\tilde{\rho},\vec{\q})$,
$q_0=q_0(\tilde{\pi},\vec{\p};\tilde{\rho},\vec{\q})$ do satisfy the
mass-shell equations (\ref{equatttion}). We add that the $c$-number
on rhs of (\ref{aallgge}) should be  proportional to the Planck
constant $\hbar$.

For classical $\kappa$-deformed fields the $c$-number on rhs of the
relation (\ref{aallgge}) vanishes  and in such a case  the relations
(\ref{aallgge}) describe  the  $\kappa$-deformed braided
oscillators. Substituting in (\ref{19}) the relation (\ref{aallgge})
with vanishing $c$-number one obtains the commutator of
$\kappa$-braided free fields, describing $\kappa$-deformation of
classical fields.
\begin{equation}
[\;{\hat \varphi}^{\rm{cl}}(x)\,,{\hat
\varphi}^{\rm{cl}}(y)\;]_{{\hat \star}_\kappa} = 0\;.
\label{braiclass}
\end{equation}

In quantum $\kappa$-deformed field theory  the nonvanishing
$c$-number on rhs of eq. (\ref{aallgge}) is proportional to Dirac
delta with the argument determined by $\kappa$-deformed
three-momentum addition law. Such a term in general case is
specified in Appendix (see also (\ref{xxy})).

The oscillators ${\hat A}(p_0\,,\vec{p})$ which are present in the
relation (\ref{aallgge})  carry the fourmomentum $p_\mu =
(p_0,\vec{p})$
\begin{equation}
P_\mu \rhd {\hat A}(p_0\,,\vec{p}) = {p}_\mu{\hat
A}(p_0\,,\vec{p})\;,\label{aacttion}
\end{equation}
restricted by the modified $\kappa$-deformed mass-shell conditions
(\ref{equatttion}). The product of two oscillators carry
respectively  the fourmomenta determined by the coproduct rule
(\ref{c4})
\begin{eqnarray}
&&P_i \rhd \left({\hat A}(p_0\,,\vec{p}){\hat A}(q_0\,,\vec{q})
\right) = \left(p_i{{\rm e}^{\frac{q_0}{2\kappa}}} + q_i{{\rm
e}^{-\frac{p_0}{2\kappa}}}\right)\left({\hat A}(p_0\,,\vec{p}){\hat
A}(q_0\,,\vec{q}) \right)\;, \label{carrying}\\
&&P_0 \rhd \left({\hat A}(p_0\,,\vec{p}){\hat A}(q_0\,,\vec{q})
\right) = \left(p_0 + q_0\right)\left({\hat A}(p_0\,,\vec{p}){\hat
A}(q_0\,,\vec{q}) \right)\;. \label{carrying1}
\end{eqnarray}
Applying the rules (\ref{carrying}),(\ref{carrying1}) to both
products of oscillators occurring in (\ref{aallgge}) we should
obtain the same eigenvalues. We get the following relations:

i) the  class of addition laws for threemomenta
\begin{equation}
\vec{p}^{\ (1+2)} = \left(\vec{\p}f + \vec{\q}g \right){\rm
e}^{q_0/2\kappa} + \left(\vec{\p}k + \vec{\q}l \right){\rm
e}^{-p_0/2\kappa} = \vec{p}^{\ (2+1)} \;,\label{additionlaw11}
\end{equation}
satisfied as identity for arbitrary values of $\vec{p}$ and
$\vec{q}$. Such additional law imply the following change of the
three-momentum Dirac delta in the oscillator algebra (see also
(\ref{jpa3}),(\ref{jpa4}) in Appendix)
\begin{equation}
\delta^{(3)}(\vec{\p}-\vec{\q})\rightarrow
\delta^{(3)}\left(\left[\vec{\p}f + \vec{\q}g \right]{\rm
e}^{q_0/2\kappa} - \left[\vec{\p}k + \vec{\q}l \right]{\rm
e}^{-p_0/2\kappa}\right)\,,\label{xxy}
\end{equation}

ii) the standard addition law for energy
\begin{equation}
{p}_0^{(1+2)} = p_0+q_0 = {p}_0^{(2+1)}\;.\label{additionlaw111}
\end{equation}
The relations (\ref{aallgge}) and (\ref{additionlaw11}) depend on
four functions ${f},\;{g},\;{k},\;{l}$; the values of $p_0$ and
$q_0$ are determined from the pair of equations (\ref{equatttion}).
%It follows from the relations (\ref{dadzbog1}) that the relation
%(\ref{additionlaw111}) is satisfied as an identity.
Different choices of $\kappa$-deformed oscillator algebras
(\ref{aallgge}) can be classified by the corresponding explicite
form of threemomentum addition laws described by
(\ref{additionlaw11}). We add that the relation (\ref{aallgge})
decomposes into four sets of bilinear relations for
$\kappa$-deformed creation and annihilation operators (see
Appendix).

In order to obtain explicit formulas we shall consider now the
particular cases of general relation (\ref{aallgge}). Firstly we
recall the algebraic scheme providing the Abelian addition law
\cite{11}, \cite{1} and further, we  present the framework providing
the $c$-number commutator in the case of the non-Abelian addition
law (\ref{csa4}) (see also \cite{12}-\cite{ny}).

i) {\it Abelian addition law} \cite{11}, \cite{1}, \cite{10}.

 In such
a case the functions occurring in (\ref{aallgge}) are the following
\begin{eqnarray}
f = {\rm
e}^{-\frac{\q_0}{2\kappa}}\;\;,\;\;g=0\;\;,\;\;k=0\;\;,\;\;l= {\rm
e}^{\frac{\p_0}{2\kappa}}\;\;;\;\;p_0=\p_0\;\;,\;\;q_0=\q_0\;.
\label{gwiazdy1}
\end{eqnarray}
The mass-shell conditions (\ref{equatttion}) take the form
\begin{equation}
 C^2_\kappa(\p_0\,,\vec{\p})-M^2 =
0\;\;\;,\;\;\;C^2_{\kappa}(\q_0\,,\vec{\q})-M^2 = 0\;,
\label{newformshell}
\end{equation}
i.e. one should put the following on-shell energy values
\begin{equation}
\p_0^{(\pm)}= \pm \omega_{\kappa}(\vec{\p})\;\;,\;\;  \q_0^{(\pm)} =
\pm \omega_{\kappa}(\vec{\q})\;. \label{uuuu}
\end{equation}
 One can
check that from (\ref{additionlaw11}) and (\ref{additionlaw111}) we
get the Abelian addition laws for
 threemomenta and energy %\cite{11}, \cite{1}
\begin{equation}
\vec{\p}^{(1+2)} = \vec{\p}+ \vec{\q}=\vec{\p}^{(2+1)}\;\;\;,\;\;\;
{\p}_0^{(1+2)} = {\p}_0+ {\q}_0={\p}_0^{(2+1)} \;, \label{gwiazdki}
\end{equation}
where $\p_0$, $\q_0$ lie on the mass-shells (\ref{newformshell}). We
point out here that in \cite{11} we have chosen the Abelian addition
law (\ref{gwiazdki}) as the selection principle for the choice of
$\kappa$-deformed statistics.

It can be shown  that for the choices given by (\ref{gwiazdy1}) the
$\kappa$-deformed oscillator algebra (\ref{aallgge}) can be written
  in the following standard classical form  $([\;A,B\;]_\circ :=
A\circ B - B\circ A)$\footnote{In fact the rhs of second relation
(\ref{sskappaccr2}) is multiplied by Planck constant $\hbar$. In our
consideration we put $\hbar =1$.}
\begin{eqnarray}
&&[\;a_\kappa({\p}),a_\kappa({\q})\;]_{\circ} =  [\;a_\kappa ^\dag
({\p}),a_\kappa^\dag ({\q})\;]_{\circ} = 0%\;,\label{sskappaccr2}%\\
\;\;,\;\; [\;a_\kappa^\dag ({\p}),a_\kappa ({\q})\;]_{\circ} =
2\Omega_\kappa(\vec{\p})\delta^{(3)} (\vec{\p}-\vec{\q})\;,~~~~~~\;~%\;,%\nonumber
\label{sskappaccr2}
\end{eqnarray}
where the creation and annihilation operators occurring in
(\ref{sskappaccr2}) are $(\p_0^{(\pm)}=\pm\omega_\kappa(\vec{\p})$;
$\p_0^{+}>0)$
\begin{eqnarray}
a_{\kappa} ({{\p}}) := {\hat A}(\p_0^{(+)},\vec{\p})\;,\;
a_\kappa^\dag ({{\p}}) := {\hat A}(\p_0^{(-)},\vec{\p})={\hat
A}(-\p_0^{(+)},-\vec{\p}) \;;\;\Omega_\kappa(\vec{\p}) \equiv\kappa
\sinh\left(\frac{\omega_{\kappa}(\vec{\p})}{\kappa}\right)\;\label{5554}
\end{eqnarray}
and the $\kappa$-deformed $\circ$-multiplication of two oscillators
was
given in \cite{11}, \cite{1}. \\
The $\kappa$-deformed multiplication $\circ$ is defined in such a
way that the following relation is valid  \cite{10}
\begin{equation}
\hat{\varphi}(x) \hat{\star}_{\kappa} \hat{\varphi}(y) =
\hat{\varphi}(x)\circ \hat{\varphi}(y)\;,\label{gwidon}
\end{equation}
where we define
\begin{eqnarray}
&&\hat{\varphi}(x)\circ \hat{\varphi}(y) = \frac{1}{(2\pi)^{3}} \int
d^4p\,d^4q\, {\rm e}^{i(p_{\mu}x^{\mu} +q_{\mu}y^{\mu})}{\hat
A}(p_0,\vec{p})\circ {\hat A}(q_0,\vec{q})
 \;\cdot~~~~~~\nonumber\\
&&~~~~~~~~~~~~~~~~~~~~~~~~~~~~~~~~~~~~~~\cdot \;
\delta(C^2_\kappa(p_0,\vec{p})-M^2)
 \delta(C^2(q_0,\vec{q})-M^2)
\;.\label{czartwor1}
\end{eqnarray}
Subsequently
\begin{equation}
[\;\hat{\varphi}(x),\hat{\varphi}(x)\;]_{\star_{\kappa}}=
[\;\hat{\varphi}(x), \hat{\varphi}(y)\;]_{\circ}\;,\label{gwidon400}
\end{equation}
what after insertion of the relations (\ref{sskappaccr2}) permits to
obtain the $c$-number value of  field commutator described by the
formulae (\ref{micro1}),(\ref{pauli}).

One can point out that in consistency with the classical fourmomenta
addition law (\ref{gwiazdki}) the choice (\ref{gwiazdy1}) is the one
which provides the particular $c$-number field commutator function
(\ref{pauli}) which is invariant under the classical fourdimensional
translations $x_\mu \to x_\mu + a_\mu$, i.e. we obtain  the
$\kappa$-deformed commutator function depending on the
four-coordinate difference $x_\mu- y_\mu$ \cite{9}.

ii) {\it Non-Abelian $\kappa$-deformed addition law and
$\kappa$-deformed flip operator.}

We choose now the arbitrary functions in (\ref{aallgge}) as follows
\begin{eqnarray}
f = 1\;\;,\;\;g=0\;\;,\;\;k=0\;\;,\;\;l= 1\;\;;\;\; p_0 =
\p_0\;\;,\;\;q_0 = \q_0\;. \label{gwiazdy1000}
\end{eqnarray}
The mass-shell conditions (\ref{equatttion}) take the form
\begin{equation}
C^2_\kappa(\p_0\,,\vec{\p}{\rm e}^{\q_0/2\kappa})-M^2 =
0\;\;\;,\;\;\;C^2_{\kappa}(\q_0\,,\vec{\q}{\rm
e}^{-\p_0/2\kappa})-M^2 = 0\;, \label{gwidon800}
\end{equation}
or more explicitly
\begin{equation}
\p_0 = \epsilon \omega_{\kappa} (\vec{\p}{\rm
e}^{\q_0/2\kappa})\;\;\;,\;\;\;\q_0=
\epsilon'\omega_{\kappa}(\vec{\q}{\rm e}^{-\p_0/2\kappa})\;.
\label{gwidon1800}
\end{equation}
Then one gets from (\ref{aallgge}) the following uniquely determined
$\kappa$-deformed oscillator algebra
\begin{eqnarray}
{\hat A}(\p_0\,,\vec{\p}) {\hat A}(\q_0\,,\vec{\q}) -
J\left(\stackrel{p',\ q'}{\p,\q}\right) {\hat
A}(\q_0\,,\vec{\q}{{\rm e}^{-\frac{\p_0}{\kappa}}}){\hat
A}(\p_0\,,\vec{\p} {{\rm e}^{\frac{\q_0}{\kappa}}}) = c-{\rm
number}\;, \label{idismatri}
\end{eqnarray}
where the $c$-number in (\ref{idismatri}) is proportional to the
following Dirac delta
\begin{equation}
\delta^{(3)}(\vec{\p}-\vec{\q})\rightarrow \delta^{(3)}(\vec{\p}{\rm
e}^{\mathcal{\q}_0/2\kappa}-\vec{\q}{\rm
e}^{-\mathcal{\p}_0/2\kappa})\,,
\end{equation}
and occurs only in creation-annihilation (or annihilation-creation)
sector (see Appendix,(\ref{jpa3}),(\ref{jpa4})).
 The formula (\ref{idismatri}) can be
written as well with the use of the $\kappa$-deformed flip operator
\begin{eqnarray}
{\hat A}(\p_0\,,\vec{\p}) {\hat A}(\q_0\,,\vec{\q}) -{\hat
\tau}_{\kappa} \left({\hat A}(\p_0\,,\vec{\p}) {\hat
A}(\q_0\,,\vec{\q})\right) = c-{\rm number}\;, \label{flip}
\end{eqnarray}
where
\begin{eqnarray}
{\hat \tau}_{\kappa} \left({\hat A}(\p_0\,,\vec{\p}) {\hat
A}(\q_0\,,\vec{\q})\right) = J\left(\stackrel{p',\ q'}{\p,\q}\right)
 {\hat A}(\q_0\,,\vec{\q}{{\rm e}^{-\frac{\p_0}{\kappa}}}){\hat A}(\p_0\,,\vec{\p}
{{\rm e}^{\frac{\q_0}{\kappa}}})\;, \label{flip1}
\end{eqnarray}
and $\p_0,\q_0$ are the solutions of the equations
(\ref{gwidon1800}). Using twice the formula (\ref{flip1}) and the
property $J\left(\stackrel{p,\ q}{\p,\q}\right)J\left(\stackrel{\p,\
\q}{p,q}\right)=1$ it is easy to see that the flip operator
$\hat{\tau}_\kappa$ satisfies the condition
\begin{eqnarray}
{\hat \tau}_{\kappa}^2 =1\;.\label{taukw}
\end{eqnarray}

In order to adjust the set of relations (\ref{jpa1})-(\ref{jpa4})
(see Appendix) to the choice (\ref{gwiazdy1000}) of functions
$f,g,k,l,p_0$ and $q_0$ we should solve the equations
(\ref{gwidon1800}). The equation for the solutions
$\p_0^{(\epsilon,\epsilon')}=\p_0^{(\epsilon,\epsilon')}(\vec{\p},\vec{\q})$
looks as follows
\begin{equation}
\p_0^{(\epsilon,\epsilon')}=\epsilon \omega_\kappa\left(\vec{\p}\,
{\rm
exp}\left[\frac{\epsilon'}{2\kappa}\omega_\kappa\left(\vec{\q}\,{\rm
exp}(-\p_0^{(\epsilon,\epsilon')}/2\kappa)\right)\right]\right)\,.
\label{masssquare}
\end{equation}
One gets the second set of energy values using the relation
\begin{equation}
\q_0^{(\epsilon,\epsilon')}(\vec{\q}\,,\vec{\p};\kappa)=
\p_0^{(\epsilon,\epsilon')}(\vec{\p}\,,\vec{\q};-\kappa)\,.
\label{bysymm}
\end{equation}
The formulae for $\p_0^{(\epsilon,\epsilon')}$ and
$\q_0^{(\epsilon,\epsilon')}$  after inserting in (\ref{idismatri})
provide explicite example of four set (\ref{jpa1})-(\ref{jpa4}) (see
Appendix) of the $\kappa$-deformed twisted oscillator algebra for
creation and annihilation operators. After inserting these algebraic
relations in field commutator (\ref{19}) we can show  that the
commutator (\ref{zap}) one obtains a $c$-number value.

\section{General algebraic structure of $\kappa$-deformed oscillator algebras}

Let us consider firstly standard undeformed theory.
 The standard algebra describing the bosonic oscillators $a(\p),a^\dag(\q)$ looks as follows
\begin{equation}
[\;a^\dag(\p)\,,a^\dag(\q)\;]=[\;a^\dag(\p)\,,a^\dag(\q)\;]=0\;\;\;,\;\;\;
[\;a^\dag(\p)\,,a^\dag(\q)\;]=2\omega(\vec{\p})\delta^3(\vec{\p}-\vec{\q})\,,
\label{fockkk}
\end{equation}
where $\p\equiv(\p_0=\omega(\vec{\p}),\vec{\p})$.

The $\kappa$-deformation of the algebra (\ref{fockkk}) can be
introduced in two ways:
\begin{itemize}

\item[i)] {\it The deformation of oscillator algebra (\ref{fockkk}) described by the
general  $\kappa$-deformed multiplication rule.}

In order to describe the general algebra (\ref{aallgge}) we
introduce new $\kappa$-deformed multiplication $\odot$
 generalizing the $\circ$-multiplication given in \cite{11}, \cite{1} as follows
\begin{eqnarray}
&&\hat{A}(\p)|_{\p_0=\epsilon\omega_\kappa(\vec{\p})}\odot \hat{A}(\q)|_{\p_0=\epsilon'\omega_\kappa(\vec{\p})}:=J\left(\stackrel{p(\p,\q),q(\p,\q)}{ \p,\q}\right) \label{newwmult}\\
&&~~~~~~\cdot{\hat
A}(p_0^{(\epsilon,\epsilon')}(\p,\q)\,,\vec{p}(\p,\q)){\hat
A}(q_0^{(\epsilon,\epsilon')}(\p,\q)\,,\vec{q}(\p,\q))|_{\p_0=\tilde{\pi},\q_0=\tilde{\rho}}
\,,\nonumber
\end{eqnarray}
where $\epsilon=\epsilon'=1$ describes the creation-creation sector,
$\epsilon=-\epsilon'$ the creation-annihilation sector, and
$\epsilon=\epsilon'=-1$ the annihilation-annihilation sector.

$\qquad$ We described in such a way the first part of binary
relation (\ref{aallgge}) depending on six arbitrary functions. One
can consider a subclass of the multiplication rule (\ref{newwmult})
which permit to describe the relation (\ref{aallgge}) as the
standard oscillator algebra (\ref{fockkk}) with the generalized
$\kappa$-deformed multiplication. For such a purpose the
multiplication (\ref{newwmult}) should satisfy additional relation
permiting as well to express second part of the binary relation
(\ref{aallgge}) by the use of multiplication rule (\ref{newwmult}),
namely
\begin{eqnarray}
&&\hat{A}(\q)|_{\q_0=\epsilon'\omega_\kappa(\vec{\q})}\odot \hat{A}(\p)|_{\p_0=\epsilon\omega_\kappa(\vec{\p})}=J\left(\stackrel{p'(\p,\q),q'(\p,\q)}{ \p,\q}\right)\label{condmult}\\
&&\qquad\qquad\qquad\qquad\qquad\qquad\cdot{\hat
A}(q_0^{(\epsilon,\epsilon')}(\p,\q)\,,\vec{q}(\p,\q){\rm e}^{-p_0^{(\epsilon,\epsilon')}(\p,\q)/\kappa}) \nonumber\\
&&\qquad\qquad\qquad\qquad\qquad\qquad\cdot{\hat
A}(p_0^{(\epsilon,\epsilon')}(\p,\q)\,,\vec{p}(\p,\q){\rm
e}^{q_0^{(\epsilon,\epsilon')}(\p,\q)/\kappa})|_{\p_0=\tilde{\pi},\q_0=\tilde{\rho}}\,.\nonumber
\end{eqnarray}
The validity of (\ref{condmult}) restricts six arbitrary functions
occurring in  (\ref{aallgge}) in the following way
\begin{equation}
p_i^T(\p,\q)\equiv p_i(\q,\p)=q_i(\p,\q){\rm
e}^{-p_0^{(\epsilon,\epsilon')}(\p,\q)/\kappa}\;\;\;,\;\;\;
(p_0^{(\epsilon,\epsilon')})^T(\p,\q)=q_0^{(\epsilon,\epsilon')}(\p,\q
)\,, \label{cona2}
\end{equation}
\begin{equation}
q_i^T(\p,\q)\equiv q_i(\q,\p)=p_i(\p,\q){\rm
e}^{q_0^{(\epsilon,\epsilon')}(\p,\q)/\kappa}\;\;\;,\;\;\;
~~(q_0^{(\epsilon,\epsilon')})^T(\p,\q)=p_0^{(\epsilon,\epsilon')}(\p,\q)\,.
\label{cona1}
\end{equation}
The algebra (\ref{aallgge}) with the choice
(\ref{cona2}),(\ref{cona1}) can be written  as follows
\begin{equation}
[\;a(\p)\,,a(\q)\;]_\odot=[\;a^\dag(\p)\,,a^\dag(\q)\;]_\odot=0\;\;\;,\;\;\;
[\;a^\dag(\p)\,,a(\q)\;]_\odot= c-{\rm number}\,. \label{fockkk2}
\end{equation}
The relations (\ref{fockkk2}) describe the class of
$\kappa$-deformed oscillator algebras with generalized
$\kappa$-deformed multiplication, parametrized by three arbitrary
functions solving  the conditions (\ref{cona2}),(\ref{cona1})

$\qquad$ In particular if $p_0=\p_0,q_0=\q_0$ and we choose (see
(\ref{gwiazdy1})
\begin{equation}
\vec{p}(\p,\q)={\rm e}^{-\frac{\q_0}{2\kappa}}\vec{\p}\;\;\;,\;\;\;
\vec{q}(\p,\q)={\rm e}^{\frac{\p_0}{2\kappa}}\vec{\q}\,,
\end{equation}
we get
\begin{equation}
\vec{p}^{\ T}(\p,\q)={\rm
e}^{-\frac{\p_0}{2\kappa}}\vec{\q}\;\;\;,\;\;\; \vec{q}^{\
T}(\p,\q)={\rm e}^{\frac{\q_0}{2\kappa}}\vec{\p}\,,
\end{equation}
where $\p_0=\p_0^{(\epsilon)}\,,\q_0=\q_0^{(\epsilon')}$. \\
 We see
that the choice (\ref{gwiazdy1}) which provides  first example of
$\kappa$-multiplication \cite{11} obviously satisfies the conditions
(\ref{cona2}),(\ref{cona1}).

\item[ii)] {\it The general $\kappa$-deformed flip operator.}

Let us describe the commutator of two standard bosonic oscillators
as follows
\begin{equation}
 [\;\hat{A}(\p)\,,\hat{A}(\q)\;]\equiv \hat{A}(\p)\hat{A}(\q)-\hat{\tau}_0[\hat{A}(\p)\hat{A}(\q)]=c-{\rm number}\,,
 \label{cococo}
\end{equation}
where $\p_0=\epsilon\omega(\vec{\p})$ and
$\hat{\tau}_0(\hat{A}\hat{B})=\hat{B}\hat{A}$. In particular if
$n=2$, the symmetrization operator $S_2$ is described in terms of
flip operator as follows
\begin{equation}
S^{(0)}_2=\frac{1}{2}(1+\hat{\tau}_0)\,.
\end{equation}
One can look for the $\kappa$-deformation of the classical flip
operator $\hat{\tau}_0$, which leads to the following
$\kappa$-deformation of the commutator (\ref{cococo})
\begin{equation}
[\;\hat{A}\,,\hat{B}\;]\rightarrow
[\;\hat{A}\,,\hat{B}\;]_{\hat{\tau}_\kappa}:=\hat{A}\hat{B}-\hat{\tau}_\kappa
\hat{A}\hat{B}\,,
\end{equation}
where we should assume that
\begin{equation}
[\;\hat{\tau}_\kappa,\Delta(P_\mu)\;]=0\,. \label{flpe}
\end{equation}

We supplement the multiplication (\ref{newwmult}) with the following
definition of $\kappa$-deformed flip operation
\begin{eqnarray}
&& \hat{\tau}_\kappa \left[\hat{A}(\p)|_{\p_0=\epsilon\omega_\kappa(\vec{\p})}\odot \hat{A}(\q)|_{\q_0=\epsilon'\omega_\kappa(\vec{\q})}\right] \label{flipgen}\\
&&\equiv\hat{\tau}_\kappa \left[J\left(\stackrel{p(\p,\q),q(\p,\q)}{
\p,\q}\right){\hat
A}(p_0^{(\epsilon,\epsilon')}(\p,\q)\,,\vec{p}(\p,\q)) {\hat
A}(q_0^{(\epsilon,\epsilon')}(\p,\q)\,,\vec{q}(\p,\q))\right]|_{\p_0=\tilde{\pi},\q_0=\tilde{\rho}}\nonumber\\
&&:=J\left(\stackrel{p'(\p,\q),q'(\p,\q)}{ \p,\q}\right) {\hat
A}(q_0^{(\epsilon,\epsilon')}(\p,\q)\,,\vec{q}(\p,\q){\rm e}^{-p_0^{(\epsilon,\epsilon')}(\p,\q)/\kappa}) \nonumber\\
&&~~~~~~~~\qquad\qquad\qquad\cdot{\hat
A}(p_0^{(\epsilon,\epsilon')}(\p,\q)\,,\vec{p}(\p,\q){\rm
e}^{q_0^{(\epsilon,\epsilon')}(\p,\q)/\kappa})|_{\p_0=\tilde{\pi},\q_0=\tilde{\rho}}\,,\nonumber
\end{eqnarray}
consistently with (\ref{flpe}). Further, one can show that for any
choice of the multiplication $\odot$ we obtain
\begin{equation}
 \hat{\tau}_\kappa^2=1\,.
 \end{equation}
 \;\;\;\;\;In particular one can consider a $\kappa$-statistics described only by $\kappa$-deformed flip operator
 with standard undeformed multiplication rule
\begin{equation}
\hat{A}(\p)\odot \hat{A}(\q)\equiv \hat{A}(\p) \hat{A}(\q)\,.
\end{equation}
Such a case  is obtained by putting $f=l=1$, $g=k=0$ and
$p_0=\p_0\,,q_0=\q_0$, and  the flip operator is uniquely defined as
follows (see (\ref{flip1}))
\begin{eqnarray}
{\hat \tau}_{\kappa} [{\hat A}(\p) {\hat A}(\q)] =
J\left(\stackrel{p',\ q'}{\p,\q}\right)
 {\hat A}(\q_0\,,\vec{\q}{{\rm e}^{-\frac{\p_0}{\kappa}}}){\hat A}(\p_0\,,\vec{\p}
{{\rm e}^{\frac{\q_0}{\kappa}}})\;. \label{flip2}
\end{eqnarray}

 \item[iii)] {\it The most general $\kappa$-deformed oscillator algebra.}

We shall consider now the most general $\kappa$-deformed oscillator
algebra (\ref{aallgge}), which for the energy-momentum dispersion
relation described by the mass-shell conditions (\ref{equatttion})
leads to $c$-number commutator function. It appears that such
algebra can be described by the composition of the multiplication
(\ref{newwmult}) and $\kappa$-deformed flip (\ref{flipgen}) in the
following way
\begin{equation}
(1-\hat{\tau}_{\kappa})\left[\hat{A}(\p)|_{\p_0=\epsilon\omega_\kappa(\vec{\p})}\odot
\hat{A}(\q)|_{\q_0=\epsilon'\omega_\kappa(\vec{\q})}\right]=c-{\rm
number}\;.
\end{equation}

If we describe the 2-particle sector of standard bosonic Hilbert
space $\mathcal{H}_2$ as symmetrized tensor product of one particle
Hilbert spaces $\mathcal{H}_1$
\begin{equation}
\mathcal{H}_2=S_2\circ(\mathcal{H}_1\otimes \mathcal{H}_1)\,,
\end{equation}
the $\kappa$-deformed multiplication modifies  the tensor product as
follows
\begin{equation}
\mathcal{H}_1\otimes \mathcal{H}_1\rightarrow
\mathcal{H}_1\otimes_\kappa \mathcal{H}_1\,,
\end{equation}
where $\otimes_\kappa$ is obtained from the  "braided
multiplication" $\odot$ of oscillators  (see (\ref{newwmult})) and
$\kappa$-deformed twist $\hat{\tau}_\kappa$ changes the
symmetrization operator
\begin{equation}
S_2\rightarrow S^\kappa_2=\frac{1}{2}(1+\hat{\tau}_\kappa)\,.
\end{equation}
\end{itemize}

In this paper we limited our considerations to the most general
binary algebraic relations.   Only in the  case of algebra
(\ref{aallgge}) with the particular choice of arbitrary functions
given by (\ref{gwiazdy1}) (see \cite{11}, \cite{1}) the products of
arbitrary number of $\kappa$-deformed creation and annihilation
operators have been introduced, and all sectors of corresponding
$\kappa$-deformed Hilbert space have been considered.

\section{Final Remarks}

In this paper we consider the general structure of $\kappa$-deformed
binary oscillator algebras and their applicability for description
of free $\kappa$-deformed quantum fields, with $c$-number commutator
functions.
 Our  considerations  generalize  recent examples of $\kappa$-deformed statistics studied by
 present authors \cite{11}, \cite{1}, \cite{10}, Arzano and Marciano \cite{12},
 Young and Zegers \cite{13}, \cite{ny} and Govindarajan et all \cite{go}.
We consider two different classes of the $\kappa$-deformed
oscillators which differ by their $\kappa$-deformed energy-momentum
relations: in Sect. 2 we assume that the oscillators
$\hat{A}(p_0,\vec{p})$ are put on standard $\kappa$-deformed mass
shell (\ref{casimir}), and in Sect. 3 we modify the standard
$\kappa$-deformed mass-shell conditions in a way providing
$c$-number field commutators.

The basic results obtained in this paper are the following:
\begin{itemize}
\item[i)] We considered in Sect. 2 the algebra of deformed oscillators  on
$\kappa$-deformed mass-shell, corresponding to noncommutative field
theory with standard multiplication of noncommutative fields, via
standard $\kappa$-star product (\ref{newstar1}). In such a case for
arbitrary choice of $\kappa$-deformed oscillator algebra the
$\kappa$-deformed quantum fields are not free. One can show that the
commutator  function is necessarily bilinear in field operators.

\item[ii)] We studied in Sect. 3 the general algebra of oscillators with modified
$\kappa$-deformed mass-shell conditions (see (\ref{gwidon800})),
corresponding to noncommutative field theory with modified
multiplication law of noncommutative fields (see (\ref{slawion1w}))
represented by nonstandard $\kappa$-star product (\ref{slawion1a}).
In such a case one obtains for large class of $\kappa$-deformed
oscillator algebras (see (\ref{aallgge}) and Appendix) the
$\kappa$-deformed free quantum fields which are characterized by the
$c$-number commutator function. It should be stressed that only in
such a case one can look for the formulation of  the
$\kappa$-deformed perturbative description of interacting
$\kappa$-deformed field theory and consider  the suitable
generalization of Feynman diagram technique.

\item[iii)] In Sect. 4 we did show that the general $\kappa$-deformed statistics
providing $\kappa$-deformed quantum fields with $c$-number commutators is obtained
by the composition of general $\kappa$-deformed multiplication and the $\kappa$-deformed flip operator.

\end{itemize}

We point out finally that in this paper we do not discuss the
$\kappa$-covariance of the $\kappa$-deformed oscillator algebras.
The results of \cite{13}, \cite{ny} suggest however that the
$\kappa$-covariance is not consistent with the relation
(\ref{dadzbog1}) which we had to postulate in order to obtain the
$c$-number $\kappa$-fields commutator. It appears therefore that the
freedom of choice present in the formula (\ref{aallgge}) might not
permit  to obtain both the manifest $\kappa$-covariance as well as
the $c$-number commutator function.

\section{Appendix: General algebra of $\kappa$-deformed creation and annihilation operators}

One can rewrite the mass-shell conditions (\ref{equatttion}) as the
set of four nonlinear algebraic equations describing eight classes
of energy-momentum dispersion relations $p_0\equiv
p_0^{(\epsilon,\epsilon')}(\tilde{\pi},\vec{\p};\tilde{\rho},\vec{\q})$,
$q_0\equiv
q_0^{(\epsilon,\epsilon')}(\tilde{\pi},\vec{\p};\tilde{\rho},\vec{\q})$
($\epsilon = \pm 1$, $\epsilon' = \pm 1$)
\begin{eqnarray}
p_0 = \epsilon \omega_{\kappa}([\vec{\p}f + \vec{\q}g]{\rm
e}^{q_0/2\kappa})\;\;\;,\;\;\; q_0= \epsilon'
\omega_{\kappa}([\vec{\p}k + \vec{\q}l]{\rm e}^{-p_0/2\kappa})\;,
\label{nowerownaniaaa}
\end{eqnarray}
where
\begin{eqnarray}
\lim_{\kappa \to \infty} p_0^{(\epsilon,\epsilon')} = \epsilon
\omega(\vec{p})\;\;\;,\;\;\; \lim_{\kappa \to
\infty}q_0^{(\epsilon,\epsilon')} = \epsilon' \omega(\vec{q})\;.
\label{nowerownaniaaa2}
\end{eqnarray}
The values $p_0^{(1,\epsilon')}>0$, $(q_0^{(\epsilon,1)}>0)$
describe the on-shell values of $p_0$ ($q_0$) which select ${\hat
A}(p_0^{(1,\epsilon')}\,,\vec{p})$, ${\hat
A}(q_0^{(\epsilon,1)}\,,\vec{q})$ as $\kappa$-deformed creation
operators; the on-shell energy values $p_0^{(-1,\epsilon')}<0$,
 $(q_0^{(\epsilon,-1)}<0)$ are required to define the $\kappa$-deformed
annihilation operators ${\hat A}(p_0^{(-1,\epsilon')}\,,\vec{p})$,
${\hat A}(q_0^{(\epsilon,-1)}\,,\vec{q})$. If we denote
\begin{eqnarray}
p^{(\epsilon,\epsilon')} = (
p_0^{(\epsilon,\epsilon')},\vec{p}(\tilde{\pi},\vec{\p};\tilde{\rho},\vec{\q}))\;\;\;,\;\;\;q^{(\epsilon,\epsilon')}
=
(q_0^{(\epsilon,\epsilon')},\vec{q}(\tilde{\pi},\vec{\p};\tilde{\rho},\vec{\q}))\;,\label{mpla}
\end{eqnarray}
one can rewrite the relations (\ref{aallgge}) as the  four set of
relations describing creation-creation, creation-annihilation,
annihilation-creation and annihilation-annihilation sectors.

If we insert in proper way the on-shell energy values
$p_0^{(\epsilon,\epsilon')}$, $q_0^{(\epsilon,\epsilon')}$ (see
(\ref{nowerownaniaaa}),(\ref{mpla})) the relations (\ref{aallgge})
decompose into the following set of relations:

i) creation-creation algebra
\begin{eqnarray}
&&J\left(\stackrel{p^{(++)},q^{(++)}}{ \p,\q}\right)
\hat{A}(p_0^{(++)}\,,\vec{\p}f + \vec{\q}g)\cdot
\hat{A}(q_0^{(++)}\vec{\p}k +
\vec{\q}l)   \label{jpa1}\\
&&\qquad=J\left(\stackrel{p'^{(++)},q'^{(++)}}{ \p,\q}\right)
\hat{A}(q_0^{(++)}\,, [\vec{\p}k + \vec{\q}l]{\rm
e}^{-p_0^{(++)}/\kappa})\cdot \hat{A}(p_0^{(++)}\,,[\vec{\p}f +
\vec{\q}g]{\rm e}^{q_0^{(++)}/\kappa})\;,\nonumber
\end{eqnarray}
ii) annihilation-annihilation algebra
\begin{eqnarray}
&&J\left(\stackrel{p^{(--)},q^{(--)}}{ \p,\q}\right)
\hat{A}(p_0^{(--)}\,,\vec{\p}f + \vec{\q}g)\cdot
\hat{A}(q_0^{(--)}\vec{\p}k +
\vec{\q}l)   \label{jpa2}\\
&&\qquad=J\left(\stackrel{p'^{(--)},q'^{(--)}}{ \p,\q}\right)
\hat{A}(q_0^{(--)}\,, [\vec{\p}k + \vec{\q}l]{\rm
e}^{-p_0^{(--)}/\kappa})\cdot \hat{A}(p_0^{(--)}\,,[\vec{\p}f +
\vec{\q}g]{\rm e}^{q_0^{(--)}/\kappa})\;,\nonumber
\end{eqnarray}
iii) creation-annihilation algebra
\begin{eqnarray}
&&J\left(\stackrel{p^{(+-)},q^{(-+)}}{ \p,\q}\right)
\hat{A}(p_0^{(+-)}\,,\vec{\p}f + \vec{\q}g)\cdot
\hat{A}(q_0^{(-+)}\vec{\p}k +
\vec{\q}l)   \label{jpa3}\\
&&\qquad-J\left(\stackrel{p'^{(+-)},q'^{(-+)}}{ \p,\q}\right)
\hat{A}(q_0^{(-+)}\,, [\vec{\p}k + \vec{\q}l]{\rm
e}^{-p_0^{(+-)}/\kappa})\cdot \hat{A}(p_0^{(+-)}\,,[\vec{\p}f +
\vec{\q}g]{\rm e}^{q_0^{(-+)}/\kappa})\;\nonumber\\
&&=2N_{\kappa}^{(+-)}(\vec{\p},\vec{\q})\,
\delta^{(3)}\left[(\vec{\p}f + \vec{\q}g){\rm e}^{q_0^{(-+)}/\kappa}
- (\vec{\p}k + \vec{\q}l){\rm e}^{-p_0^{(+-)}/\kappa}\right]
\;,\nonumber
\end{eqnarray}
iv) annihilation-creation algebra
\begin{eqnarray}
&&J\left(\stackrel{p^{(-+)},q^{(+-)}}{ \p,\q}\right)
\hat{A}(p_0^{(-+)}\,,\vec{\p}f + \vec{\q}g)\cdot
\hat{A}(q_0^{(+-)}\vec{\p}k +
\vec{\q}l)   \label{jpa4}\\
&&\qquad-J\left(\stackrel{p'^{(-+)},q'^{(+-)}}{ \p,\q}\right)
\hat{A}(q_0^{(+-)}\,, [\vec{\p}k + \vec{\q}l]{\rm
e}^{-p_0^{(-+)}/\kappa})\cdot \hat{A}(p_0^{(-+)}\,,[\vec{\p}f +
\vec{\q}g]{\rm e}^{q_0^{(+-)}/\kappa})\;\nonumber\\
&&=2N_{\kappa}^{(-+)}(\vec{\p},\vec{\q})\,
\delta^{(3)}\left[(\vec{\p}f + \vec{\q}g){\rm e}^{q_0^{(+-)}/\kappa}
- (\vec{\p}k + \vec{\q}l){\rm e}^{-p_0^{(-+)}/\kappa}\right]
\;.\nonumber
\end{eqnarray}
The threemomentum Dirac delta in (\ref{jpa3}),(\ref{jpa4}) describes
in accordance with the formula (\ref{additionlaw11}) the
$\kappa$-deformed  threemomentum conservation law for the process of
creation/annihilation of field quanta with the threemomentum
$(\vec{\p}f + \vec{\q}g)$ and annihilation of the quanta with the
momentum $(\vec{\p}k + \vec{\q}l)$.

\section*{Acknowledgments}
This paper has been financially supported by Polish Ministry of
Science and Higher Education grant NN202318534.

\end{document}